\documentstyle[twocolumn,prl,aps,psfig]{revtex}        

\begin{document}
\twocolumn[%
\hsize\textwidth\columnwidth\hsize\csname@twocolumnfalse\endcsname
\draft

\title{Interacting particles at a metal-insulator transition}

\author{ Cosima Schuster$^{1}$, Rudolf A.\ R\"{o}mer$^{2}$, and
  Michael Schreiber$^2$\\ $^{1}$Institut f\"{u}r Physik,
  Universit\"{a}t Augsburg, D-86135 Augsburg, Germany\\ $^2$Institut
  f\"{u}r Physik, Technische Universit\"{a}t, D-09107 Chemnitz,
  Germany}

\date{$Revision: 1.13 $, compiled \today}

\maketitle

\begin{abstract}
  We study the influence of many-particle interaction in a system
  which, in the single particle case, exhibits a metal-insulator
  transition induced by a finite amount of onsite pontential
  fluctuations. Thereby, we consider the problem of interacting
  particles in the one-dimensional quasiperiodic Aubry-Andr\'{e}
  chain.
  We employ the density-matrix renormalization scheme to investigate
  the finite particle density situation.  In the case of
  incommensurate densities, the expected transition from the
  single-particle analysis is reproduced.  Generally speaking,
  interaction does not alter the incommensurate transition.
  For commensurate densities, we map out the entire phase diagram and
  find that the transition into a metallic state occurs for attractive
  interactions and infinite small fluctuations --- in contrast to the
  case of incommensurate densities.  Our results for commensurate
  densities also show agreement with a recent analytic renormalization
  group approach.
\end{abstract}

] 
\narrowtext \tighten

\section{Introduction}
\label{sec-intro}

The metal-insulator transition (MIT) in disordered electronic systems
has been the subject of intense research activities over the last two
decades and still continues to attract much attention. For
non-interacting electrons in disordered systems \cite{KraM93} the
scaling hypothesis of localization \cite{AbrALR79} can successfully
predict many of the universal features of the MIT. However, the
influence of many-particle interactions on the MIT is not equally well
understood \cite{BelK94} and recent investigations of an apparent MIT
in two-dimensional systems even question the main assumptions of the
scaling hypothesis \cite{KraDS96,BelK98,PudBPB97}.
In the single-particle case, the one-dimensional quasiperiodic
Aubry-Andr\'{e} model is known rigorously to exhibit an MIT for all
states in the spectrum as a function of the quasiperiodic potential
strength $\mu$ \cite{AubA80}.  The ground state wave function is
extended for $\mu < 1$ and localized for $\mu >1$. The system at
$\mu_{\text{c}}=1$ is critical: there the wave functions decrease
algebraically, not exponentially as in the localized case.  This
behavior is contrary to the localization of non-interacting particles
in a one-dimensional random potential, where the ground state wave
function becomes localized for infinite small disorder strength.
An ingeneous theoretical approach to the interplay of interactions and
disorder is based on the two-interacting-particles (TIP) problem in
one-dimensional random potential \cite{She94,She96a,RomS97a,FraMPW95}.
Furthermore, numerical results for spinless fermions in a random
potential at finite particle density have given additional insight
\cite{SchSSS98,SchJWP98}. In general, these investigations have shown
that changes in the wave function interferences due to many-particle
interactions \cite{RomSV99,RomSV00} can lead to a rather large
enhancement of the localization lengths in one and two dimensions
\cite{SchJWP98,LeaRS99,RomLS99b}.
Recently, we examined the TIP problem in a quasiperiodic potential by
means of the transfer-matrix and the decimation method together with a
careful finite-size-scaling analysis \cite{EilGRS99,EilRS01}.  We
found that it exhibits the MIT at $\mu_{\text{c}}=1$ as in the
single-particle case --- independent of interaction.

As an independent extension of these low-density TIP results, Chaves
and Satija \cite{ChaS97} have studied a model of nearest-neighbor
interacting spinless fermions \cite{YanY66a,YanY66b,KorBI93} at finite
particle density in the same quasiperiodic potential by means of
Lanczos diagonalization for small systems up to chain length $M=13$.
They find evidence for a critical region in which the behavior of the
charge stiffness \cite{ShaS90,SutS90,RomP95} is different from its
behavior in the metallic and localized regimes. In order to reach much
larger system sizes for interacting systems, we employ the numerical
density-matrix renormalization group (DMRG) \cite{Whi93} which has
been shown to be very useful \cite{Whi98}. In particular, the ground
state properties of interacting many-particle systems in one dimension
can be obtained very accurately \cite{PaiPR97,Sch99}. In the present
paper, we shall study the quasiperiodic model of Ref.\ \cite{ChaS97}
at various densities and interaction strengths by DMRG and compare our
results to the TIP data.

For the interacting many-body system in a quasiperiodic potential we
recover the transition at $\mu_{\text{c}}=1$ independent of interaction,
provided we consider densities like $\rho=1/2$ which are
incommensurate compared to the wave vector of the quasiperiodic
potential --- an irrational multiple of $\pi$.  Thus, the low-density
TIP case is comparable to finite but incommensurate densities.  On the
other hand, for commensurate densities, we find that the system can be
completely localized even for $\mu\ll 1$, due to a Peierls resonance
between electronic and quasiperiodic potential degrees of freedom.
Whereas for repulsive interactions the ground state remains localized,
we find a region of extended states for attractive interaction due to
the interplay between interaction and quasiperiodic potential.  The
behavior may be described by a weak-coupling renormalization group
(RG) treatment \cite{VidMG99}.  Thus, the physics of the model is
dominated by whether the density is commensurate or incommensurate and
only then by interaction effects.

The paper is organized as follows.  The finite-density many-body
system is introduced in section \ref{sec-finite-density}. Results
for incommensurate and commensurate densities are presented in section
\ref{sec-inc-dens} and \ref{sec-finite-density-results}, respectively.
We summarize and conclude in section \ref{sec-concl}.

\section{The nearest-neighbor Hamiltonian at finite density and the DMRG}
\label{sec-finite-density}

Let us consider $N$ interacting spinless fermions on a ring of
circumference $M$ in the Aubry-Andr\'{e} potential \cite{AubA80} such that
\begin{eqnarray}
  H & = & -t\sum_m\left( c^{\dag}_{m+1}c^{}_m + {\rm h. \ c.}\right)+
  \nonumber \\ & & \mbox{ } V\sum_m n_{m+1}n_m +2\mu\sum_m n_m
  \cos{(\alpha m + \beta)}.
\label{eq-ham-finite-density}
\end{eqnarray}
The operators $c_m^{\dag}$, $c_m$, and $n_m$, respectively, denote as
usual Fermi creation, annihilation and number operators; $2 \mu \cos(
\alpha m + \beta)$ is the quasiperiodic potential of strength $\mu$
with $\alpha /2 \pi$ being an irrational number.  $\beta$ is an
arbitrary phase shift and we choose $\alpha/2 \pi= (\sqrt{5} -1)/2$,
i.e., the inverse of the golden mean. 
In
addition, we set $t=1$. The particle density is $\rho=N/M$. 
The model has been studied extensively in two limits, namely,
independent particles \cite{Koh83}, $V=0$, and the clean case, $\mu=0$
\cite{YanY66a,YanY66b,KorBI93}.  The first case corresponds of course
to the single-particle problem discussed briefly in section
\ref{sec-intro}. In the second case, the model can be mapped onto the
anisotropic Heisenberg (XXZ) model for which a closed Bethe-ansatz
solution exists \cite{YanY66a,YanY66b,KorBI93}. It shows three
distinct phases at zero temperature. For half filling ($\rho=1/2$) and
strong repulsive interaction ($V>2$) the system is a
charge-density-wave insulator.  For weak and intermediate interaction
strength and away from half filling it is a metal and can be described
as a Luttinger liquid with linear energy dispersion and gapless
excitations \cite{ShaS90,Hal81a}.  The metal, at least, is separated
for all fillings by a first-order transition at $V=-2$ from an
insulator where the fermions form clusters (phase-separated state),
corresponding to the ferromagnetic state of the spin model
\cite{KorBI93}.

In a previous analysis \cite{ChaS97} of the interacting quasiperiodic
model (\ref{eq-ham-finite-density}), a large enhancement of the Drude
weight $D$ (or Kohn stiffness) \cite{Koh64} and the superconducting
fluctuations in the ground state was found near the first-order
transition at $V\approx -2$ using exact Lanczos diagonalization.
However, the system sizes attainable by the diagonalization were
restricted to $M\le 13$.
Using the DMRG, it is possible to extend the tractable system lengths
to about $M\approx 100-200$.  We use the finite lattice algorithm for
non-reflectionsymmetric models as described in Ref.\ \cite{Sch96}. In
our simulations we perform five lattice sweeps and keep 300 (small
systems) to 500 (larger systems, $M=$89 and 144) states per block.  As
our observable of the phase transition we choose the phase sensitivity
\begin{equation}
  M\Delta E=M(-1)^{N}[E(0)-E(\pi)]
\label{eq-phase-sensitivity}
\end{equation}
of the ground state \cite{Sch99}, which is connected to $D$ in the
clean case (Luttinger phase) and equal to $\pi^2 D$ for
non-interacting fermions.  Here $E(\Phi)$ measures the reaction of the
system due to a twist in the boundary condition,
$c_{M+1}=\exp(i\Phi)c_1$ \cite{ShaS90,SutS90}. The prefactor
$(-1)^{N}$ cancels the odd-even effects \cite{Los92}.  In addition, it
is believed that the phase sensitivity matches the character of the
wave function. It does not depend on system size if the wave function
is extended, e.g., for $\mu=0$, and it decreases exponentially for
large systems if the wave function is localized \cite{Sch99}. As
argued in Ref.\ \cite{ChaS97} this can be transferred to the critical
case. Therefore, we suppose that the phase sensitivity decreases
algebraically with increasing system size, if the wave function is
critical. Thus, we have to compare at the very least three different
chain lengths in order to characterize the length dependence of the
wave function.

\section{Incommensurate particle densities}
\label{sec-inc-dens}

In order to systematically study finite-size effects, we should in
principle compute $M\Delta E$ for fixed particle density $\rho$ and
increasing $M$.  However, with fixed $\alpha/2 \pi= (\sqrt{5} -1)/2$,
the quasi-periodic potential is incompatible with periodic boundary
conditions and the results for $M\Delta E$ depend on the choice of
$\beta$. Averaging over many potentials with different $\beta$ can in
principle be used in order to obtain $\beta$-independent results.
However, we have found that the statistical fluctuations for densities
like $\rho=1/2$, $1/3$, $1/4$ at fixed $\alpha/2\pi=(\sqrt{5} -1)/2$
are rather large such that we could not detect any significant $M$
dependence of the phase sensitivity.

As is customary in the context of quasiperiodic systems \cite{Gri99},
the value of $\alpha/2 \pi$ may be approximated by the ratio of
successive Fibonacci numbers --- $F_n=F_{n-2}+F_{n-1}= 0$, $1$, $2$,
$3$, $5$, $8$, $13$, $\ldots$.  In this way, choosing
$\alpha/(2\pi)=F_{n-1}/F_n$ and $M=F_n$, we can retain the periodicity
of the quasiperiodic potential on the ring.  Averaging over different
$\beta$ values is then no longer necessary since the computed
ground-state energy of the finite-density Hamiltonian does not depend
on $\beta$.

For a convenient comparison with results obtained for the disordered
systems \cite{SchSSS98,SchJWP98}, we would now want to study the phase
sensitivity at various densities, say, $\rho=1/2$.  However, we then
have to use $M=34$, $144$, $610$, $\ldots$. For third filling, we
would be restricted to $M=21$, $144$, $987$ and so on.  In principle,
such large system sizes $M \gg 100$ can be treated within the DMRG
\cite{Sch96}, but the necessary accuracy ($\approx 10^{-6}$) is very
hard to obtain for periodic boundary conditions.  In addition, if the
phase sensitivity decreases with system size, there is no possibility
to decide whether the decrease is algebraic or exponential because the
obtained value of $M\Delta E$ for such large system sizes is already
zero within the computational accuracy. 

\begin{figure}
  \centerline{\psfig{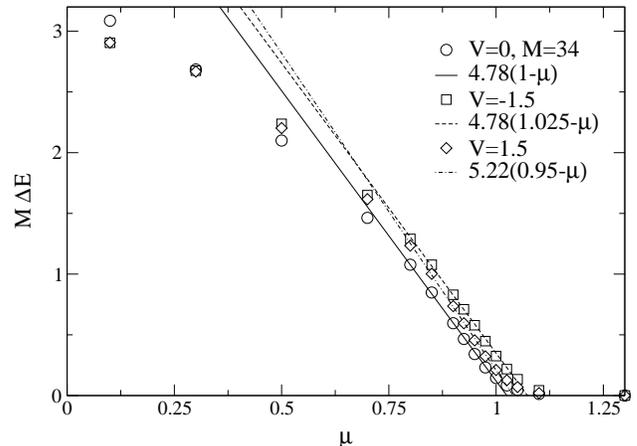}}
\caption{
  Phase sensitivity versus $\mu$ at fixed system size $M=34$ for
  $V=-1.5$ (squares), $0$ (circles), $1.5$ (diamonds) at
  (incommensurate) half filling.  The lines are fits corresponding
  $M\Delta E\sim |\mu-\mu_{\text{c}}|$.}
\label{fig-fixedM}
\end{figure}

For example, let us comment on $\rho=1/2$.  First, we check the
occurrence of an MIT near $\mu=1$ at fixed $M$.  Calculating the phase
sensitivity at fixed $M$ for different values of $\mu$ and $V$, we
find that there is a clear transition near $\mu_{\text{c}}=1$, as seen
in Fig.\  \ref{fig-fixedM}. 
Whereas for small $\mu$ the phase sensitivity for non-interacting
particles is larger than for interacting --- as it is in the clean
case ---, interaction slightly enhances the phase sensitivity in the
extended regime for intermediate $0.5\lesssim \mu \lesssim 1$.
Near $\mu=1$, it seems that the MIT is not shifted appreciably within
the accuracy of our calculation. As shown in Fig.\ \ref{fig-fixedM},
fitting the data results in a power-law behavior close to $\mu=1$ with
slightly larger $\mu_{\text{c}}=1.025$ for attractive interaction and
slightly smaller $\mu_{\text{c}}=0.95$ for repulsive. Also, close to
the transition, the phase sensitivity can be described by $M \Delta E
\propto (\mu_{\text{c}} - \mu)^{\nu}$ with $\nu\approx 1$. This is
similar to the TIP situation where $\nu=1$ was found for the behavior
of the finite-size-scaled localization lengths
\cite{EilGRS99,EilRS01}.

\begin{figure}
  \centerline{\psfig{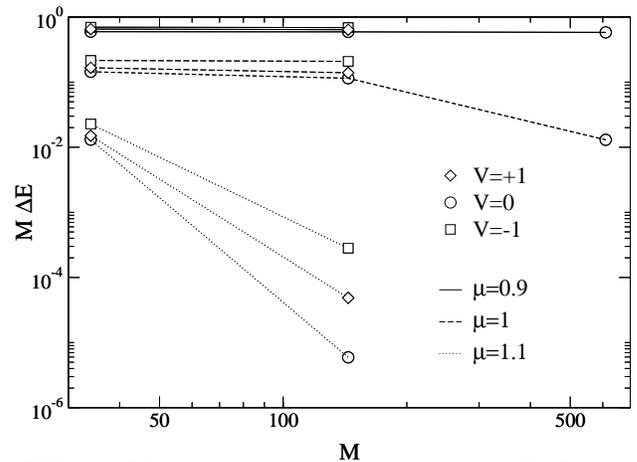}}
\caption{
  Phase sensitivity versus system size for $V=-1$ (squares), $0$
  (circles), $1$ (diamonds) at (incommensurate) half filling.  The
  three cases $\mu=0.9$ (solid lines) , $\mu=1$ (dashed lines), and
  $\mu=1.1$ (dotted lines) are compared. }
\label{fig-dmrg-phase}
\end{figure}

In the non-interacting case, we can also study the length dependence
of the phase sensitivity by a standard diagonalization routine. We
investigate the system sizes $M=34$, $144$, and $610$.  $M=8$ is
excluded because $\alpha/(2\pi)=5/8=0.625$ differs too much from the
true value $(\sqrt{5}-1)/2\approx 0.618$. As shown in Fig.\ 
\ref{fig-dmrg-phase}, the phase sensitivity at $V=0$ is 
different in the localized, critical and extended regimes.  Thus, for
a system of free fermions at incommensurate density, we reproduce the
expected transition at $\mu_{\text{c}}=1$ in agreement with Refs.\ 
\cite{EilGRS99}.  Similar plots can be made for attractive and
repulsive interactions at $\rho=1/2$ as also shown in Fig.\ 
\ref{fig-dmrg-phase}. Unfortunately only the two system sizes $M=34$
and $144$ are available due to numerical instabilities for the
interacting system. Therefore, further conclusions appear rather
speculative.  Nevertheless, we conjecture that the MIT at
$\mu_{\text{c}}\approx1$ for incommensurate densities is only weakly influenced
by attractive and repulsive interaction in agreement with the results
obtained in the TIP case \cite{EilGRS99,EilRS01}.

\section{Commensurate particle densities}
\label{sec-finite-density-results}
For the reasons outlined in the last section, we shall now turn our
attention to the behavior at the commensurate densities $\rho_i
\approx \lim_{n\rightarrow\infty} F_{n-i}/F_n \approx 0.618$, $0.382$,
$0.236$, and $0.146$ corresponding to $i=1, \ldots,4$ in the
following, where $\alpha=2\pi\rho_1$.

\subsection{The Peierls-like metal-insulator-transition}
\label{sec-peierls}

Let us first comment on the densities $\rho_1$ and $\rho_2$ and note
that the case of $\rho_1$ is identical to $\rho_2$ due to the
particle-hole symmetry in Eq.\ (\ref{eq-ham-finite-density}).
Furthermore, the electronic Fermi wave-vector is $k_{\text{F}}= \pi
\rho_1$ and therefore we have the {\em resonance condition}
$\alpha=2k_{\text{F}}$ for the wave-vector $\alpha$ of the
quasiperiodic potential.  Consequently, we expect {\em critical
  behavior} as found for the periodic potential at the Peierls
transition \cite{Pei55}, i.e., a transition from insulating to
metallic phase at $V= -\sqrt{2}$ and infinite small potential
strength. Peierls \cite{Pei55} considered about 50 years ago the
behavior of one-dimensional tight-binding electrons due to a periodic
lattice distortion and found the same resonance condition.  We
therefore call the transition observed in the present quasiperiodic
system a {\em Peierls-like} transition. The critical behavior for
fermions in a periodic potential is equivalent to the critical
behavior of the periodic hopping model. There the site-dependent
hopping amplitude is given by $t_m=1-\delta\cos{q m}$.  In this
so-called Su-Schrieffer-Heeger model \cite{SuSH80} the MIT occurs at
$V=-\sqrt{2}$ and $\delta\to 0$, if filling factor $\rho$ and wave
vector $q$ of the periodic hopping are commensurate, i.e.,
$\rho=q/(2\pi)$ or $q = 2k_{\text{F}}$ \cite{SchE98}. This happens for
$\rho=1/2$ and dimerization, one-third filling and trimerization, and
so on \cite{Sch99}. For increasing $\delta$, $V_{\text{MIT}}$ decreases.
The quasiperiodic model is expected to show this transition, at
densities $\rho_1$ and $\rho_2$, because $\alpha = 2k_{\text{F}}$ in
these cases. We have checked numerically that the Peierls-like
transition is indeed observed in these cases.

\begin{figure}
  \centerline{\psfig{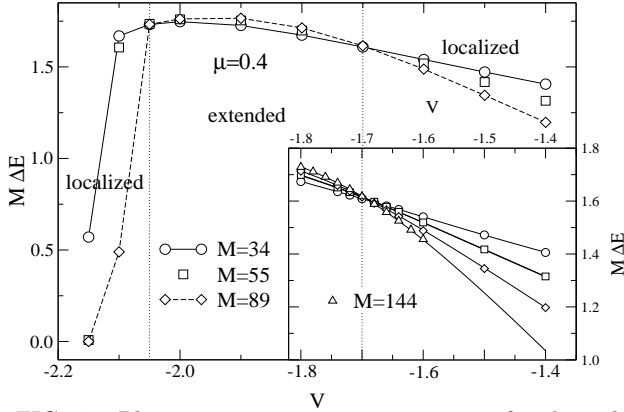}}
\caption{
  Phase sensitivity versus interaction for three different system
  sizes, density $\rho_3$ and $\mu=0.4$. The lines for $M=34$ and $89$
  are guides to the eye, highlighting the different
  finite-size-scaling behaviors in the various regimes. The dotted
  lines indicate the transition between the localized and the extended
  wave functions. Inset: Additional data close to the transition and
  $M=144$ ($\triangle$) included. The lines denote the curves
  constructed by non-linear finite-size scaling.}
\label{fig-dmrg-phase-interaction}
\end{figure}

Even for the other commensurate densities $\rho_3$, $\rho_4$,
$\ldots$, where the resonance condition is not strictly fulfilled,
the critical behavior is still expected to be similar to the
Peierls-case as shown in Ref.\ \cite{VidMG99}. In our case, if $\mu$
is larger than a certain minimal $\mu_{\text{min}}$ value, we find that
the phase diagrams for $\rho_3$ and $\rho_4$ are dominated by
localized states as shown in Fig.\ \ref{fig-phase-diagram}.  This
localized regime extends further on to larger interactions, $-1.4
\lesssim V < \infty$, for $\mu>0$. However, there also exists a
sizeable region of extended states starting roughly at $V=-1.4$ for
small $\mu$ and extending until $V\approx -2.5$ for larger $\mu$.
As an example how well the transition point is defined numerically
consider Fig.\ \ref{fig-dmrg-phase-interaction}, where we plot the
phase sensitivity versus interaction strength.  The localization for
$V>-1.7$ --- where $M\Delta E$ decreases with system size --- is
caused by the quasiperiodic potential and for $V<-2$ by the clustering
of the fermions in the phase-separated state.  In between, the
quasiperiodic potential is irrelevant and the ground-state
wave-function is extended as for $\mu=0$ --- $M\Delta E$ increases
with system size.

\begin{figure}
  \psfig{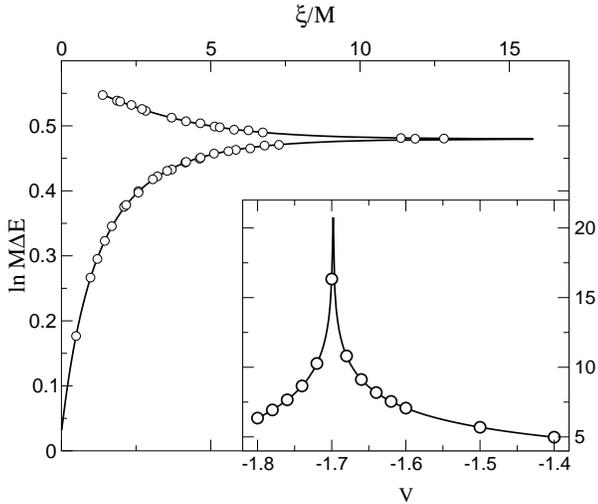}
\caption{
  Scaling function (solid line) and scaled data points  in the interval
  $V\in[-1.8,-1.4]$ close to the transition for $\mu=0.4$
  and system sizes $M= 34, 55, 89, 144$. Inset: Scaling parameter
  $\xi$ as a function of $V$.}
\label{fig-scal-mu04}
\end{figure}

\begin{figure}
  \psfig{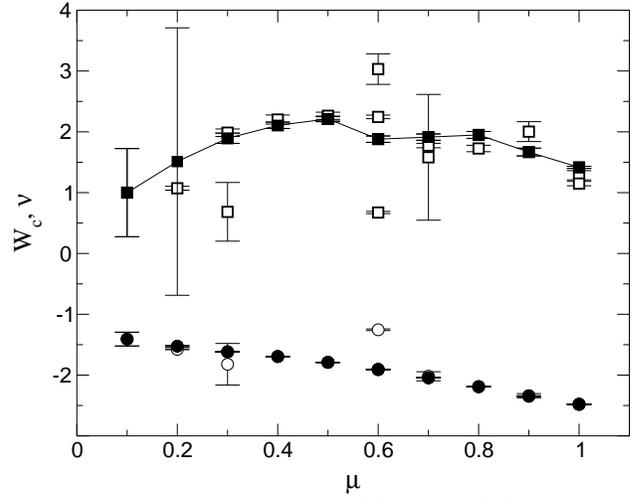}
\caption{
  Estimates of $V_{\text{c}}$ ($\circ$) and $\nu$ ($\Box$) obtained for
  $\rho_3$ by non-linear finite-size scaling at various $\mu$. The
  open symbols denote various fit functions and initial parameters,
  whereas the closed symbols indicate the best (according to
  $\chi^2$-statistics) of all these fit for a given value of $\mu$.
  Error bars mark the errors resulting from the Levenberg-Marquardt
  fitting method. }
\label{fig-wcnu-mu-all}
\end{figure}

In the inset of Fig.\ \ref{fig-dmrg-phase-interaction} we show the
transition at $\mu=0.4$ in detail.  A similar behavior close to the
transition has also been found in data obtained recently for the TIP
case \cite{EilGRS99,EilRS01}, albeit for varying $\mu$ and constant
interaction and not for varying $V$ with constant $\mu$.  We may
nevertheless perform a finite-size-scaling analysis. A simple ansatz
assuming a power-law divergence of the correlation length $\xi\sim
|V-V_{\text{c}}|^{-\nu}$ at the transition yields reasonable scaling results as
shown in Fig.\ \ref{fig-scal-mu04}.  Taking into account
non-linearities in the finite-size-scaling ansatz as in Ref.\ 
\cite{EilRS01}, we can determine $V_{\text{c}}$ and $\nu$ for various values of
$\mu$ as shown in Fig.\ \ref{fig-wcnu-mu-all}. Our results indicate
that $\nu$ rises from $\approx 1$ at $\mu=0$ to values close at $2$ at
$\mu=0.5$ and then drops back to the expected $\nu=1$ at the
Aubry-Andr\'e transition at $\mu_{\text{c}}$. We emphasize that our data, due
to the limited system sizes available, are not sufficient to
distinguish between the power-law behavior of a second order phase
transition and the expected Kosterlitz-Thouless behavior \cite{Loi99}
at a Peierls-like transition.

Thus, for repulsive and {\em weak} attractive interactions ---
including the non-interacting case ---, we find that the ground state
for $\mu > \mu_{\text{min}}$ is localized. This is in agreement with
previous studies for disordered and periodically disturbed systems
\cite{SchSSS98,Sch99,SchE98}. We emphasize that an increase of the
localization lengths as predicted by the arguments for TIP
\cite{She94} is most likely too small \cite{SchJWP98} to be detected
by the present accuracy.  For more strongly attractive interactions at
$V \approx -1.4$, the situation is more interesting.  For all
densities $\rho_i$ and $\mu>\mu_{\text{min}}$, the system shows {\em
  Peierls-like} behavior, i.e., a transition from insulating to
metallic phase at $V\approx -\sqrt{2}$.  Changing the density, the
transition occurs for decreasing particle density at increasing $\mu$.
This observation is confirmed by an investigation of the energy
spectrum reported in Ref.\ \cite{GeiKR91}.  In the non-interacting
case it contains no gap at $\mu=0$ but $M-1$ gaps at $\mu_{\text{c}}=1$ for the
given $\alpha$ \cite{AbaTW98}.  As seen in Ref.\ \cite{GeiKR91} most
of the gaps open successively for increasing $\mu$. Especially, the
first gap at $k=\alpha/2$ opens for $\mu\to 0$.  Tuning the filling
factor, the Fermi points fall into the additional gaps in the
spectrum, leading to insulating behavior.

\subsection{Comparison with the renormalization group treatment}
\label{sec-RG}

\begin{figure}
  \centerline{\psfig{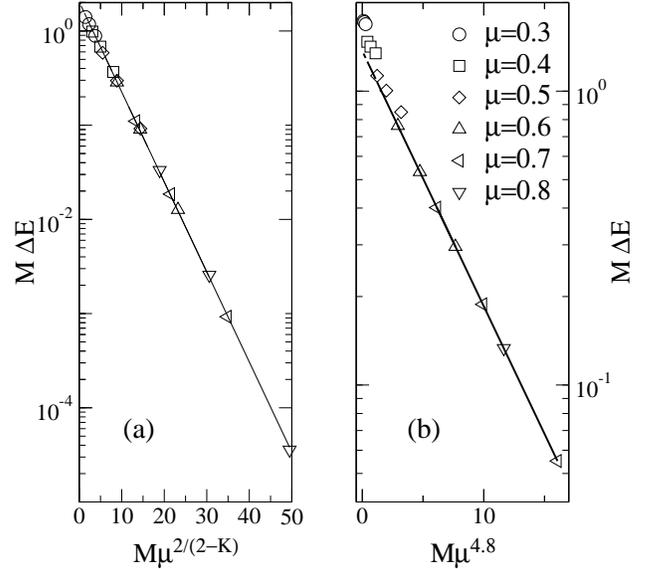}}
\caption{
  (a) Phase sensitivity versus scaled system size $M\mu^{2/(2-K)}$,
  where $M=34$, $55$, and $89$, for $V=-0.6$, filling $\rho=\rho_3$,
  and various potential strengths. The line indicates a plot of Eq.\ 
  (\protect\ref{eq-RGscaling}) with $K=\pi/2 \arccos(0.3)\approx
  1.24$. (b) Phase sensitivity versus scaled system size, $M=34$,
  $55$, and $89$, for $V=-1.5$ and $\rho_3$. The straight line is a
  fit to the data for $\mu=0.7$ and $0.8$.}
\label{fig-dmrg-logphase}
\end{figure}

In addition, the behavior seen numerically is in agreement with the
weak-coupling RG treatment \cite{VidMG99} of spinless fermions on a
Fibonacci lattice. The relevant RG equation shows the same critical
behavior as for the Peierls model, namely
\begin{equation}
\frac{dim {\rm e}^{-M/\xi}\mu}{d \ln M}=(2-K)\mu,
\end{equation}
where $K= \pi/2\arccos{(-V/2)}$ is the Luttinger parameter for half
filling. Since the Luttinger parameter does not depend strongly on the
filling, it is valid to use this analytical expression for other
fillings, too.  In accordance with this RG equation we expect, that
the phase sensitivity --- as in a disordered system \cite{SchSSS98,RunZ94}
--- decreases in the localized regime such that
\begin{equation}\label{eq-RGscaling}
  M\Delta E \sim {\rm e}^{-M/\xi} \quad {\rm with} \quad \xi^{-1} \sim
  \mu^{2/(2-K)},
\end{equation}
where it is assumed that the localization length $\xi$ is the only
relevant length scale.  We emphasize that this scaling is only
reasonable in the localized regime.  It is clearly found in the
numerical data for, e.g., $\rho_3$ and $V=-0.6$ as shown in Fig.\ 
\ref{fig-dmrg-logphase}a). Near the phase transition, it is
numerically difficult to distinguish between localized (exponential
decay with system size) and critical (algebraic decay with system
size) phases.  For small system sizes and small $\mu$, the decrease in
the phase sensitivity is always algebraic, according to the RG
equation.  The exponential decrease with $\xi$ sets in far from the
transition points as shown in Fig.\ \ref{fig-dmrg-logphase}b).  There,
the exponential scaling is found for $\mu\geq 0.5$.  We note, that
despite the straight line was obtained by a fit only to the data for
$\mu=0.7$ and $\mu=0.8$ the data for $\mu=0.5$, and $\mu=0.6$ fall
likewise on this line.  To show the deviations from this straight line
for small $\mu$ clearly, which is the most important feature in this
plot, we do not show the data for $\mu=0.8$.  We call this
intermediate region, where no universal algebraic or exponential
decrease is found, {\em transition region}.  A finite-size scaling
corresponding to the previous section near the transition with
$\xi\sim|\mu-\mu_{\text{c}}|^{\nu}$
does not lead to conclusive results.

\subsection{Enhancement near the first-order transition}
\label{sec-SL}
\begin{figure}
  \psfig{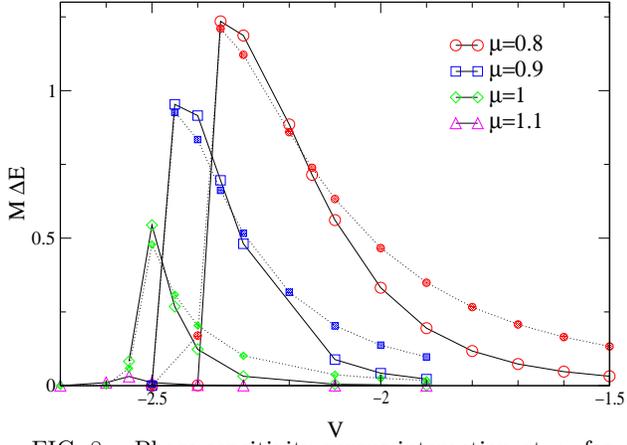}
\caption{
  Phase sensitivity versus interaction at $\rho_3$ for different $\mu$
  at size $M=34$ (small grey symbols) and $M=55$ (large open symbols).
  We compare two system sizes to indicate the regimes with the
  localized and extended ground state wave function.  The lines
  connect the data points for different $\mu$ and $M$ values.}
\label{fig-peak}
\end{figure}

As already mentioned in \cite{ChaS97}, the phase sensitivity shows a
large enhancement around $V=-2$.  We find that the first order phase
transition at $V=-2$ remains unaffected for small $\mu\lesssim 0.2$.
However, for larger $\mu$ values, the metallic phase extends towards
more negative $V$ values. For $\mu \gtrsim 0.8$ at $V_p\lesssim -2.2$
the phase sensitivity shows an unexpected sharp maximum in the
delocalized regime, as shown in Fig.\ \ref{fig-peak}.  A similar
maximum is also seen for $\mu>1$ and small system sizes ($M=34$,
$55$), but the system is insulating. The exponential scaling in the
localized region as shown in Fig.\ \ref{fig-dmrg-logphase} does not
change for $V>-2$ and $V<V_p<-2$ in the case of large $\mu$.
Therefore, we conclude that this regime belongs to the localized
regime of the quasiperiodic potential. On the other hand, no scaling
behavior is found for $V<-2$ and small and intermediate $\mu$, lower
left corner in the phase diagrams, Fig.\ \ref{fig-phase-diagram}.
Here, the localization is due to the formation of particle clusters in
the phase-separated state.  In the crossover regime between these two
localized phases a metallic state is recovered.  The maximum in the
insulating regime is a rest of this crossover.  At intermediate and
strong $\mu$ and $V<V_p<-2$ the phase sensitivity drops very rapidly
to zero within the numerical accuracy.  This behavior is comparable to
the behavior at $V=-2$ and small $\mu$ as seen in Fig.\ 
\ref{fig-dmrg-phase-interaction} at the first order transition.  Thus,
it seems that the first order transition is moved to smaller $V$ with
increasing $\mu$.  The subtle interplay between the clustering and the
quasiperiodic potential seems to delocalize the ground-state
wave-function in a small parameter region.  However, it is numerically
difficult to follow the border between these two completely different
localized phases for stronger $\mu$ and $-2<V<-3$ in our approach.

\subsection{Phase diagram}

In Fig.\ \ref{fig-phase-diagram}, we summarize our results by showing
two phase diagrams of model (\ref{eq-ham-finite-density}) for varying
quasiperiodic potential strength $\mu$ and nearest-neighbor
interaction $V$. They were obtained by studying the system size
behavior of $M\Delta E$ up to $M=144$ for $\rho_3=0.236$ and
$\rho_4=0.146$.  The localized parts of the phase diagrams for weak
attractive and repulsive interactions are not shown, because they have
no additional structure.  For $\rho_4$, the intermediate phase is
difficult to detect and is not separately marked in Fig.\ 
\ref{fig-phase-diagram}.  There we can only distinguish whether the
phase sensitivity decreases or not.  As clearly seen,
$\mu_{\text{min}}=0$ for $\rho_3$ and $\mu_{\text{min}}=0.1$ for
$\rho_4$.  In summary, the delocalized phase becomes larger for weak
$\mu$ and smaller for strong $\mu$ with decreasing density.  First,
$\mu_{\text{min}}(\rho_4)>\mu_{\text{min}}(\rho_3)$ and second, the
  crossover region beyond $V=-2$ becomes smaller and extends only to
  $\mu=0.9$.

\section{Conclusions}
\label{sec-concl}

In this work, we have studied the influence of interactions on an
metal insulator transition in one dimension using the DMRG method.
Thereby, we have compared the interplay of disorder and interactions
for the quantum system (\ref{eq-ham-finite-density}) in a
quasiperiodic potential at incommensurate and commensurate densities.
Our results suggest that the delocalization found previously for low
density TIP in the localized phase cannot simply be extrapolated to
the finite-density situation.  In the case of incommensurate densities
with nearest-neighbor (repulsive or attractive) interaction the metal
insulator transition is found at $\mu_{\text{c}}\approx 1$.  The critical
exponent, $\nu\approx 1$, is in agreement with the results of
transfer-matrix-method calculations and finite-size scaling
\cite{EilGRS99}.  This indicates that the influence of interaction at
this particular type of metal insulator transition is not strong
enough to change the universality class of the Aubry-Andr\'e model.

For commensurate densities $\rho_3$ and $\rho_4$, we have deduced the
phase diagrams of system for varying quasiperiodic potential strength
$\mu$ and interaction $V$. The numerically accessible filling factors
are commensurate with the quasiperiodic potential and yield a
Peierls-like behavior of the system with a metal insulator transition
at attractive interaction and small quasiperiodic potential strength.
Thus, although we find a strong enhancement of the phase sensitivity
at the first order transition, the physics of the model at
commensurate densities is dominated by the Peierls resonance condition
which becomes irrelevant only for strong attractive interaction.  A
remainder of the single-particle localization is still found in this
case --- the region of the phase diagram, where the ground-state
wave-function is extended, extends at most to $\mu=1$.  In addition,
we have found a change in the exponent of the localization length from
$\nu\approx 2$ to $\nu\approx 1$ at $\mu=1$.  In summary, we have seen
that the single-particle localization in the Aubry-Andr\'e model is
not influenced by interaction.  At commensurate densities, other
effects such as a Peierls-like commensurability become important and
dominate the transport properties.

\acknowledgements
We gratefully acknowledge various discussions with A.\ Eilmes
regarding the TIP results. RAR thanks the Department of Physics,
University of Ruhuna, Matara, Sri Lanka for hospitality during an
extended stay where part of the work was carried out. Financial
support is provided by the Deutsche Forschungsgemeinschaft within
Sonderforschungsbereich~393 (RAR and MS) and 484 (CS).

%
%


\begin{thebibliography}{10}

\bibitem{KraM93}
B. Kramer and A. MacKinnon, Rep. Prog. Phys. {\bf 56},  1469  (1993).

\bibitem{AbrALR79}
E. Abrahams, P.~W. Anderson, D.~C. Licciardello, and T.~V. Ramakrishnan, Phys.
  Rev. Lett. {\bf 42},  673  (1979).

\bibitem{BelK94}
D. Belitz and T.~R. Kirkpatrick, Rev. Mod. Phys. {\bf 66},  261  (1994).

\bibitem{KraDS96}
S.~V. Kravchenko {\it et~al.}, Phys. Rev. Lett. {\bf 77},  4938  (1996).

\bibitem{BelK98}
D. Belitz and T.~R. Kirkpatrick, Phys. Rev. B {\bf 58},  8214  (1998).

\bibitem{PudBPB97}
V.~M. Pudalov, G. Brunthaler, A. Prinz, and G. Bauer, {Pis'ma} ZhETF {\bf 65},
  887  (1997), [JETP Lett. 65 (1997)], cond-mat/9707054.

\bibitem{AubA80}
S. Aubry and G. {Andr\'e}, Ann. Israel Phys. Soc. {\bf 3},  133  (1980).

\bibitem{She94}
D.~L. Shepelyansky, Phys. Rev. Lett. {\bf 73},  2607  (1994).

\bibitem{She96a}
D.~L. Shepelyansky,  in {\em Correlated fermions and transport in mesoscopic
  systems}, edited by T. Martin, G. Montambaux, and J.~T.~T. {V\^{a}n}
  (Editions Frontieres, Proc. XXXI Moriond Workshop, Gif-sur-Yvette, 1996), p.\
  201.

\bibitem{RomS97a}
R.~A. {R\"{o}mer} and M. Schreiber, Phys. Rev. Lett. {\bf 78},  515  (1997).

\bibitem{FraMPW95}
K. Frahm, A. {M\"{u}ller-Groeling}, J.~L. Pichard, and D. Weinmann, Europhys.
  Lett. {\bf 31},  169  (1995).

\bibitem{SchSSS98}
P. Schmitteckert {\it et~al.}, Phys. Rev. Lett. {\bf 80},  560  (1998).

\bibitem{SchJWP98}
P. Schmitteckert, R. Jalabert, D. Weinmann, and J.-L. Pichard, Phys. Rev. Lett.
  {\bf 81},  2308  (1998).

\bibitem{RomSV99}
R.~A. {R\"{o}mer}, M. Schreiber, and T. Vojta, phys. stat. sol. (b) {\bf 211},
  681  (1999).

\bibitem{RomSV00}
R.~A. {R\"{o}mer}, M. Schreiber, and T. Vojta, Physica E  (2000).

\bibitem{LeaRS99}
M. Leadbeater, R.~A. {R\"{o}mer}, and M. Schreiber, Eur. Phys. J. B {\bf 8},
  643  (1999).

\bibitem{RomLS99b}
R.~A. {R\"{o}mer}, M. Leadbeater, and M. Schreiber, Ann. Phys. (Leipzig) {\bf
  8},  675  (1999).

\bibitem{EilGRS99}
A. Eilmes, U. Grimm, R.~A. {R\"{o}mer}, and M. Schreiber, Eur. Phys. J. B {\bf
  8},  547  (1999).

\bibitem{EilRS01}
A. Eilmes, R.~A. {R\"{o}mer}, and M. Schreiber,   (2001), {ArXiv}:
  cond-mat/0106603.

\bibitem{ChaS97}
J.~C. Chaves and I.~I. Satija, Phys. Rev. B {\bf 55},  14076  (1997).

\bibitem{YanY66a}
C.~N. Yang and C.~P. Yang, Phys. Rev. {\bf 150},  321  (1966).

\bibitem{YanY66b}
C.~N. Yang and C.~P. Yang, Phys. Rev. {\bf 150},  327  (1966).

\bibitem{KorBI93}
V.~E. Korepin, N.~M. Bogoliubov, and A.~G. Izergin, {\em Quantum Inverse
  Scattering Method and Correlation Functions} (Cambridge University Press, New
  York, 1993).

\bibitem{ShaS90}
B.~S. Shastry and B. Sutherland, Phys. Rev. Lett. {\bf 65},  243  (1990).

\bibitem{SutS90}
B. Sutherland and B.~S. Shastry, Phys. Rev. Lett. {\bf 65},  1833  (1990).

\bibitem{RomP95}
R.~A. {R\"{o}mer} and A. Punnoose, Phys. Rev. B {\bf 52},  14809  (1995).

\bibitem{Whi93}
S.~R. White, Phys. Rev. Lett. {\bf 69},  2863  (1993).

\bibitem{Whi98}
S.~R. White, Phys. Rep. {\bf 301},  187  (1998).

\bibitem{PaiPR97}
R. Pai, A. Punnoose, and R.~A. {R\"{o}mer}, preprint series of the SFB 393
  97-12, TU Chemnitz (unpublished).

\bibitem{Sch99}
C. Schuster, Ph.D. thesis, ({Shaker} {Verlag},
  {Aachen}, 1999).

\bibitem{VidMG99}
C. Vidal, D. Mouhanna, and T. Giamarchi, Phys. Rev. Lett. {\bf 83},  3908
  (1999).

\bibitem{Koh83}
M. Kohmoto, Phys. Rev. Lett. {\bf 51},  1198  (1983).

\bibitem{Hal81a}
F.~D.~M. Haldane, Phys. Rev. Lett. {\bf 47},  1840  (1981).

\bibitem{Koh64}
W. Kohn, Phys. Rev. {\bf 133},  A171  (1964).

\bibitem{Sch96}
P. Schmitteckert, Ph.D. thesis, {Universit\"{a}t} Augsburg, 1996.

\bibitem{Los92}
D. Loss, Phys. Rev. Lett. {\bf 69},  343  (1992).

\bibitem{Gri99}
U. Grimm,   (1999), {Habilitationsschrift}, {Technische} {Universit\"{a}t}
  {Chemnitz}.

\bibitem{Pei55}
R.~E. Peierls, {\em Quantum Theory of Solids} (Oxford University Press, Oxford,
  1955).

\bibitem{SuSH80}
W.~P. Su, J.~P. Schrieffer, and A.~J. Heeger, Phys. Rev. B {\bf 22},  2099
  (1980).

\bibitem{SchE98}
C. Schuster and U. Eckern, Eur. Phys. J. B {\bf 5},  395  (1998).

\bibitem{Loi99}
D. Loison, Journal of Physics: Condensed Matter {\bf 11},  L401  (1999).


\bibitem{GeiKR91}
T. Geisel, R. Ketzmerick, and G. Petschel, Phys. Rev. Lett. {\bf 66},  1651
  (1991).


\bibitem{AbaTW98}
A.~G. Abanov, J.~C. Talstra, and P.~B. Wiegman, Phys. Rev. Lett. {\bf 81},
  2112  (1998).

\bibitem{RunZ94}
K.~J. Runge and G.~T. Zimanyi, Phys. Rev. B {\bf 49}, 15212 (1994).



\end{thebibliography}

%
%

\widetext

\narrowtext

%
%

\newcommand{\figwidth}{0.95\columnwidth}

\vspace*{1ex}

\vspace*{1ex}

\vspace*{1ex}
\begin{figure}
  \psfig{figure=fig-phase.eps,width=\figwidth}
\caption{
  Phase diagrams of the system described by Eq.\ 
  (\ref{eq-ham-finite-density}) in terms of quasiperiodic potential
  strength $\mu$ and interaction $V$ for $\rho=\rho_3$ (left side) and
  $\rho=\rho_4$ (right side). An extended ground state wave function
  is marked by $\bullet$, a localized by $\diamond$. In addition, the
  transition regime (see text) is marked with a shaded $\Box$ in the
  left figure. The solid lines indicate the first-order transition at
  $V=-2$ and $\mu=0$.}
\label{fig-phase-diagram}
\end{figure}

\vspace*{1ex}

\vspace*{1ex}


\vspace*{2ex}

\vspace*{1ex}

\vspace*{1ex}

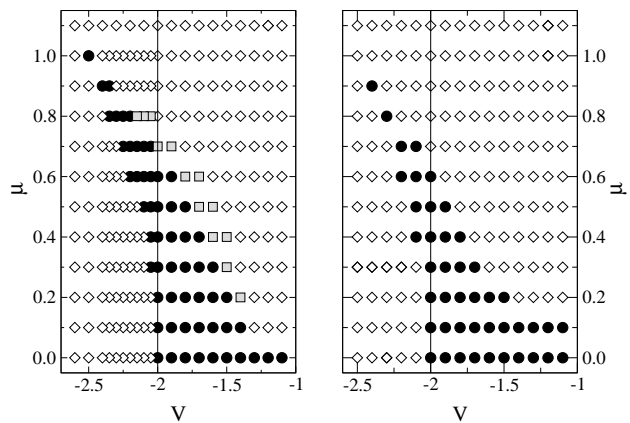
\end{document}